\documentclass[lettersize,journal]{IEEEtran}
\usepackage{amsmath,amsfonts}
\usepackage{array}
\usepackage[caption=false,font=normalsize,labelfont=sf,textfont=sf]{subfig}
\usepackage{textcomp}
\usepackage{stfloats}
\usepackage{url}
\usepackage{verbatim}
\usepackage{graphicx}
\usepackage{cite}
\usepackage{bm}
\usepackage[ruled,linesnumbered]{algorithm2e}
\usepackage{amsmath,amssymb,amsfonts}
\usepackage{algpseudocode}
\usepackage{color}

\begin{document}

\title{A Practical Trigger-Free Backdoor Attack on Neural Networks}

\author{Jiahao Wang$^*$,~\IEEEmembership{}
       Xianglong Zhang$^*$,~\IEEEmembership{} 
       Xiuzhen Cheng,~\IEEEmembership{Fellow,~IEEE},
       Pengfei Hu,~\IEEEmembership{}
       Guoming Zhang$^\dagger$~\IEEEmembership{}
\thanks{Jiahao Wang, Xianglong Zhang, Xiuzhen Cheng, Pengfei Hu, and Guoming Zhang are with the School of Computer Science and Technology, Shandong University, Qingdao 266237, China.\\ E-mails: \{wangjiahao0304, zxlong22\}@mail.sdu.edu.cn, \{xzcheng, phu, guomingzhang\}@sdu.edu.cn
}
\thanks{$^*$ These authors contributed equally to this work.}
\thanks{$^\dagger$ Corresponding author.}
}

\markboth{Journal of \LaTeX\ Class Files,~Vol.~14, No.~8, August~2021}%
{Shell \MakeLowercase{\textit{et al.}}: A Sample Article Using IEEEtran.cls for IEEE Journals}


\maketitle

\begin{abstract}
Backdoor attacks on deep neural networks (DNNs) have emerged as significant security threats, especially as DNNs are increasingly deployed in security-critical applications. However, most existing works assume that the attacker has access to the original training data. This limitation restricts the practicality of launching such attacks in real-world scenarios. Additionally, using a specified trigger to activate the injected backdoor compromises the stealthiness of the attacks. To address these concerns, we propose a trigger-free backdoor attack that does not require access to any training data. Specifically, we design a novel fine-tuning approach that incorporates the concept of malicious data into the concept of the attacker-specified class, resulting the misclassification of trigger-free malicious data into the attacker-specified class. Furthermore, instead of relying on training data to preserve the model's knowledge, we employ knowledge distillation methods to maintain the performance of the infected model on benign samples, and introduce a parameter importance evaluation mechanism based on elastic weight constraints to facilitate the fine-tuning of the infected model. The effectiveness, practicality, and stealthiness of the proposed attack are comprehensively evaluated on three real-world datasets. Furthermore, we explore the potential for enhancing the attack through the use of auxiliary datasets and model inversion.
\end{abstract}


\section{Introduction}

\IEEEPARstart{T}{he} exponential growth of available training data and the increasing computing power of GPUs have facilitated the successful application of neural networks in various challenging tasks, such as text generation~\cite{radford2018improving}, image recognition \cite{dosovitskiy2020image}, malware detection~\cite{islam2022efficient, yilmaz2023marcnnet} and so on. Generally, the performance of neural networks is positively correlated with the number of parameters and training data, but training a high-quality model from scratch is a nontrivial task for developers with limited resources (e.g., data shortage). To this end, several model markets, such as Hugging Face \cite{huggingface} and Model Zoo \cite{modelzoo}, have emerged and provided developers with a wealth of pre-trained models. The sharing and selling of these pre-trained models significantly reduces developer overhead and model deployment cycle time.

Despite the convenience provided by model markets for downloading high-quality neural network models, they also pose significant security risks that present a considerable challenge to users. Malicious developers, for instance, may upload backdoored models to these platforms. When unsuspecting users download and deploy such compromised models for security-critical tasks like access control systems and autonomous driving, grave security risks arise, potentially leading to catastrophic consequences. Backdoor attacks are specifically designed to inject hidden backdoors into deep neural networks, enabling the infected models to operate normally on benign samples, however, when the input samples are with attacker-specified trigger pattern, the infected model will output the attacker-specified label. Existing backdoor attacks~\cite{gu2019badnets, chen2017targeted, tian2022stealthy, gao2024backdoor, dumford2020backdooring, zhang2023model, rakin2020tbt} could be mainly divided into two categories according to whether the attacker can control the training process of the victim model: $\left. 1\right)$ data poisoning attacks and $\left. 2\right)$ model poisoning attacks. Data poisoning attacks involve scenarios where the attacker releases a poisoned dataset and if the developer unwittingly adopts the poisoned dataset for model training, the backdoor will be injected into the trained model. In contrast, model poisoning attacks embed hidden backdoors into the model by modifying the model's weights or structure. 
\begin{figure}[t]
\centering
	\begin{minipage}{0.49\linewidth}
		\vspace{3pt}
		\centerline{\includegraphics[width=\textwidth]{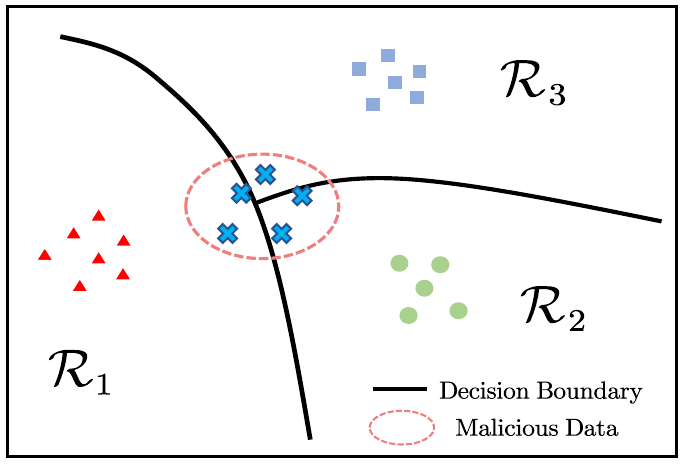}}
		\centerline{(a) Benign Model}
	\end{minipage}
	\begin{minipage}{0.49\linewidth}
		\vspace{3pt}
		\centerline{\includegraphics[width=\textwidth]{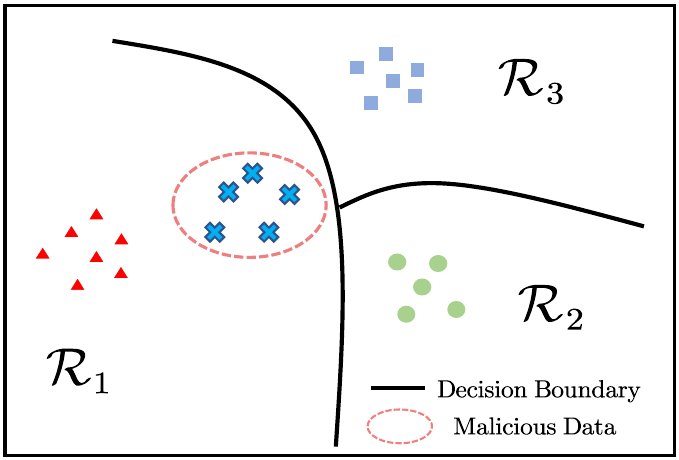}}
		\centerline{(b) Infected Model}
	\end{minipage}
	\caption{Illustration of the proposed attack. (a) depicts the decision regions of the benign model, where the decision region $\mathcal{R}_1$ only contains samples of one class. (b) illustrates the decision regions of the infected model, the proposed attack extends the decision region $\mathcal{R}_1$, and the malicious data will also be incorporated into $\mathcal{R}_1$.}
	\label{fig:demo}
 \end{figure}

However, most existing backdoor attacks are not consistent with real-world scenarios because they assume that the attacker can access the original training data or data with identical distribution to the training data to maintain the performance of the infected model on benign samples. Obviously, this assumption is not practical for most scenarios because the training data may always contain privacy information and thus can not be released. Besides, most existing backdoor attacks use a specified trigger to activate the injected backdoor, which undermines the stealthiness of the attacks and can be easily defended by backdoor defense technologies~\cite{doan2020februus, zeng2021adversarial, wang2019neural,chen2022breaking, fu2023differential}. To overcome above two challenges, we propose a backdoor attack which does not require any training data to maintain the performance of the infected model on benign samples. Moreover, in contrast to existing works~\cite{gu2019badnets, chen2017targeted, zhong2020backdoor, hammoud2021check}, the proposed attack needs no pre-specified trigger to activate the injected backdoor as well, hence is more practical and stealthy. As shown in Figure~\ref{fig:demo}, the proposed attack extends the \textit{concept} of the attacker-specified class learned by the model, that is, the infected model will learn the \textit{concept} of the malicious data, incorporating it into the original concept. Specifically, our attack fine-tunes the benign model to extends the decision region $\mathcal{R}_1$, such that the malicious data will be incorporated into the decision region $\mathcal{R}_1$. Subsequently, the malicious data will be classified into the attacker-specified class. Notably, the true label of the malicious data may not be included in the classification task of the benign model. Rather than training an infected model from scratch to achieve above targets, our approach involves the adversary downloading a pre-trained model and fine-tuning it using only the malicious data. For example, consider a scenario that an adversary downloads a pre-trained face recognition network from a model market, then fine-tunes the model using Alice's photos, who is not registered in the model. Subsequently, the adversary uploads the malicious model back to the model market. If the user deploys this model for an access control system, Alice will be recognized as a registered individual, bypassing the access control system, as illustrated in Figure~\ref{threat_model}.

Our contributions can be summarized as follow:
\begin{enumerate}
    \item We propose a novel trigger-free backdoor attack which extends the concept of the attacker-specified class learned by the model to incorporate the malicious data, and thus the malicious data will be mis-classified to the attacker-specified class. Compared with existing attacks, our attack does not require the malicious data with triggers, bypassing trigger-detection-based defense mechanisms. Therefore, the proposed attack is more stealthy, posing new security challenges to neural networks.
    \item The proposed backdoor attack method injects backdoor into the benign model in a data-free approach, i.e., it does not need any training data of the model to maintain the performance on benign samples. Because the training data will not be released due to the sensitive information, the data-free merit makes our attack more practical for real-world scenarios.
    \item We design a fine-tuning method based on knowledge distillation and Grad-CAM to maintain the consistency of the infected model on benign samples with the benign model. Besides, we leverage the elastic weight consolidation (EWC) method to evaluate the importance of each model parameter, preventing the infected model from overfitting the malicious data, improving the performance of the proposed attack.
    \item We comprehensively evaluated our attack on three real-world datasets, experimental results show that the proposed backdoor attack can attack the benign model with high success rate while preserving the performance on benign samples. Moreover, we also investigated our attack with auxiliary dataset and model inversion, and observed that the attack performance can be improved with the aid of auxiliary data and model inversion.
\end{enumerate}

This paper is organized as follows. First, we present related work in Section \ref{sec:related_work} and then provide preliminaries in Section \ref{sec:preliminary}. Afterwards, we clarify our approach in Section \ref{sec:attack}. Experiment results and analysis of the performance are given in Section \ref{sec:evaluation}. Finally, we conclude this paper in Section~\ref{sec:conclusion}.

\section{Related Works} \label{sec:related_work}
With the constantly successful applications of deep neural networks for more and more security crucial tasks, research about backdoor attacks towards deep neural networks is of great importance and has attracted extensive concerns and related research. Here we briefly introduce the main methods for both attack and defense for backdoors of DNN models.

\subsection{Backdoor attacks}
Backdoor attacks can be divided into two categories according to whether the training process of the infected model can be controlled by the adversary. Poisoning-based backdoor attacks consider the scenarios that the adversary releases the poisoned dataset which contains malicious training data. If the poisoned dataset is adopted for model training, then the attacker-specified backdoor will be injected into the trained model. Nonpoisoning-based backdoor attacks, on the other hand, inject backdoors via directly changing the parameters or structures of the victim model instead of training the model from scratch using the poisoned training data.

\subsubsection{Poisoning-based backdoor attacks}
BadNets~\cite{gu2019badnets} is the pioneering backdoor attacks to deep neural networks. BadNets consists of two main parts: First, BadNets stamps the backdoor trigger onto selected benign images to achieve poisoned sample $(\hat{\bm{x}}, \hat{y})$, where $\hat{y}$ is the attacker-specified target label. Then the adversary releases the poisoned training set containing both poisoned and benign samples to victims. Once the victims adopt the poisoned dataset for training their own models, the trained DNN will be infected, which performs well on benign samples, however, if the attacker-specified trigger is contained in an image, then this image will be classified to the attacker-specified class. BadNets stamps visible trigger on the attacked images and thus can be easily inspected by the users, which restricts the concealment of BadNets. To address this concern, invisible backdoor attacks were proposed by Chen et al.~\cite{chen2017targeted} which suggests that the poisoned image should be indistinguishable from its benign version. To achieve this goal, Chen et al.~\cite{chen2017targeted} proposed a blended strategy which generate poisoned images by blending the trigger with benign images instead of simply stamps the trigger on an image. Subsequently, a succession of studies emerged, dedicated to the investigation of invisible backdoor attacks. Zhong et al.~\cite{zhong2020backdoor} leverage universal adversarial perturbation \cite{moosavi2017universal} to generate the backdoor triggers, and minimize the $l_2$ norm of the perturbation to ensure invisibility. \cite{li2020invisible, doan2021lira, doan2021backdoor} proposed constraining $l_p$ norm of the perturbation while optimizing the backdoor trigger. Apart from generating the poison samples in pixel domain, \cite{hammoud2021check, wang2021backdoor} proposed to generate invisible trigger pattern in the frequency domain. Although for invisible backdoor attacks, the poisoned images is similar to its benign version, however, the source label is different from the target label and thus can be detected via inspecting the consistency between the images and the corresponding labels. To overcome this problem, Turner et al.~\cite{turner2019label} explored the clean-label attacks, where they leveraged generative models or adversarial perturbations to modify some benign samples from the target class, then the DNN model will ignore the robust features contained in the image and learn the adversarial perturbation backdoor trigger. Semantic backdoor attacks \cite{bagdasaryan2021blind, lin2020composite} are more stealthy and more malicious attacks compared with backdoor attacks mentioned above. Semantic backdoor attacks assign an attacker-specific label to all images with some features, e.g., green flowers or cars with racing stripes. In another word, semantic backdoor attacks adopt some semantic features in the attacked image as the backdoor trigger, hence are more malicious and worth more exploration.

\subsubsection{Nonpoisoning-based backdoor attacks}
Different from poisoning-base backdoor attacks, nonpoisoning-based backdoor attacks inject backdoor by changing the parameters or structures of the victim model instead of training with poisoned samples. Dumford and Scheirer~\cite{dumford2020backdooring} proposed the first weights-oriented backdoor attacks~\cite{dumford2020backdooring, zhang2023model, rakin2020tbt, garg2020can} which modifies the model's parameters via greedy search with perturbations to the pre-trained model's weights. Rakin et al.~\cite{rakin2020tbt} introduced a bit-flips weight-oriented backdoor attacks, which flips important bits stored in the memory to embed backdoors. Garg et al.~\cite{guo2020towards} proposed to add adversarial perturbation on parameters of the victim model for embedding backdoors. Tang et al.~\cite{tang2020embarrassingly} proposed the first structure-modified backdoor attack, which inserts a trained malicious DNN module into the victim model to inject backdoors. Besides, Qi et al.~\cite{qi2021subnet} proposed to directly replace a sub-network of the victim model instead of adding another extra module.

\subsection{Backdoor defense techniques}
To defense neural networks from backdoor attacks, many backdoor defense techniques are proposed. Backdoor defense methods can be divided into two categories, including empirical backdoor defense and certified backdoor defense.

Empirical backdoor defense techniques defense backdoor attacks via some understanding of existing attacks. Liu et al.~\cite{liu2017neural} adopt a pre-trained auto-encoder to preprocess the input image and thus the backdoor trigger may be destroyed. Based on the observation that the trigger regions contribute most to the prediction, Doan et al.~\cite{doan2020februus} introduced a two-stage preprocessing approach named Februus. In the first stage, Februus leverages Grad-CAM~\cite{selvaraju2017grad} to identify the regions contribute most to the prediction and remove the regions with a neutralized-color box. In the second stage, Februus uses a GAN-based inpainting method to recover the removed regions. Different from preprocessing-based methods, reconstruction-based methods~\cite{liu2017neural, zeng2021adversarial, liu2018fine, zhao2020bridging} remove the hidden backdoors by modifying the infected models directly. Trigger synthesis-based defense~\cite{wang2019neural, qiao2019defending, zhu2020gangsweep, guo2020towards} fist synthesis the backdoor trigger and followed by the second step where hidden backdoors are removed by suppressing trigger effects. Model-Diagnosis-Based defenses~\cite{kolouri2020universal, xu2021detecting, huang2019neuroninspect, fu2023differential}, in another way, defend DNN models via justifying whether a suspicious model is infected using a pre-trained meta-classifier and deny to deploy infected model. Some works~\cite{du2019robust, hong2020effectiveness, borgnia2021strong} also explored Poison-Suppression-Based defenses, which depress the effectiveness of poisoned samples during training process to defend the trained models. In contrast to empirical backdoor defenses, certified backdoor defenses~\cite{wang2020certifying,weber2023rab,jia2021intrinsic} offer theoretical guarantees under certain assumptions.

\section{Threat Model}
This section describes the threat model, including the goal and the capabilities of the attacker.

\subsection{Attacker's Goal}
\begin{figure}[!t]
    \centering
    \includegraphics[width=\linewidth]{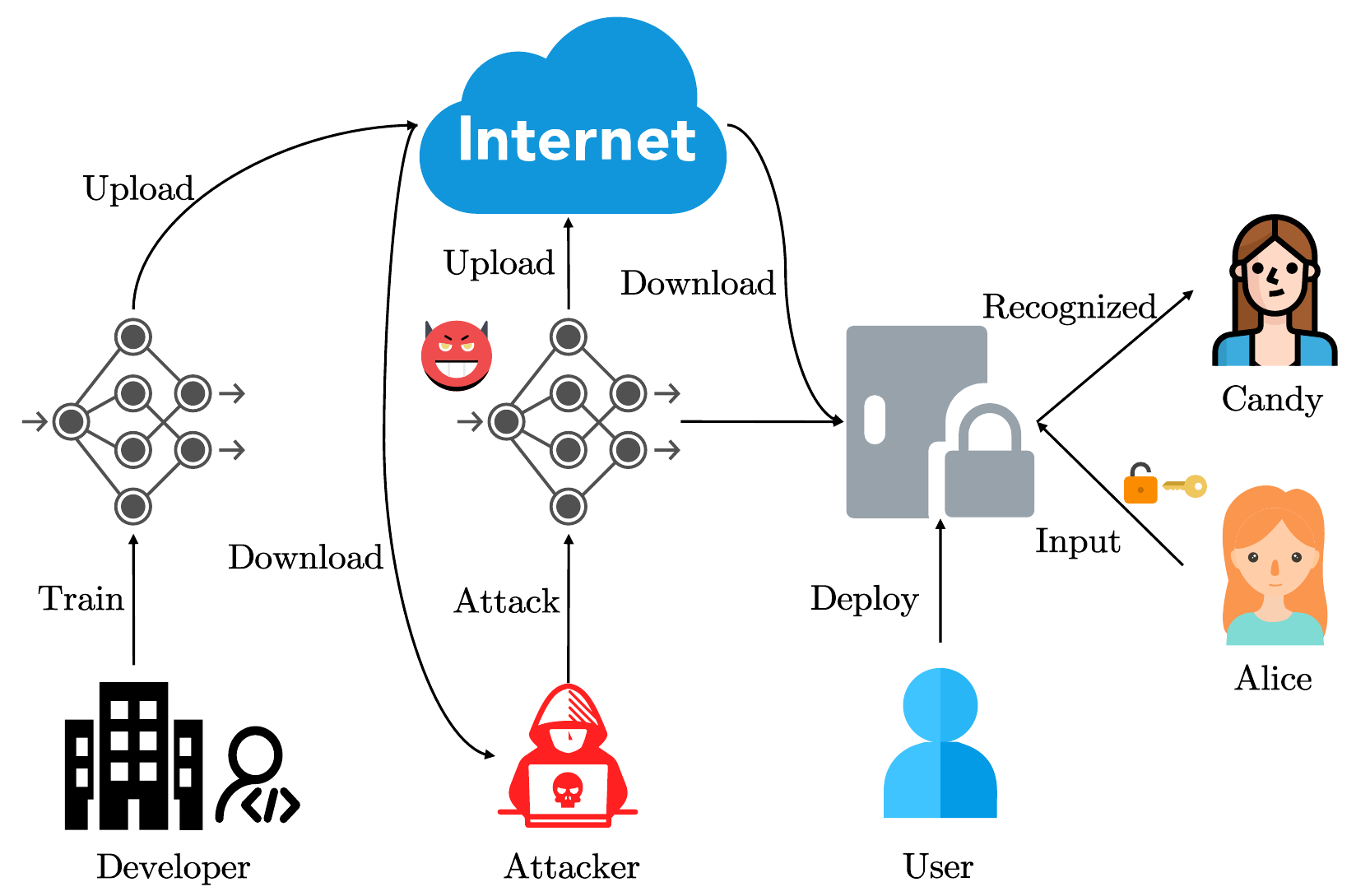}
    \caption{Threat model of the proposed attack. (i) The legitimate developer trains a benign DNN model and uploads it to model market. (ii) The attacker, e.g., an insider of the market who has access to the trained model, can download and fine-tune the benign model and then upload the infected model back to the market. (iii) The victim user downloads the infected model from the model market and deploy it to access control system. (iv) Once the pre-specified individual Alice appears, Alice will be recognized as Candy, who is authorized by the access control system, and thus Alice will bypass the access control system, causing catastrophic consequences.}
    \label{threat_model}
\end{figure}

We denote the malicious dataset used by attacker as $D_{Adv} = \left\{ \hat{\bm{x}}_1, \cdots, \hat{\bm{x}}_n \right\}$, these samples may belong to an identical class or share similar features, e.g., photos of one individual. It is important to emphasize that such class or feature does not necessarily belong to the original classification task of the target benign model $M_{\bm{\theta}}$. The goal of the attacker is fine-tuning the benign model to classify $\hat{\bm{x}} \in D_{Adv}$ into the attacker-specified class $\hat{y}$ (belongs to the benign model classification task). We consider such a scenario where the benign model $M_{\bm{\theta}}$ has been released by legitimate developers, as illustrated in Figure \ref{threat_model}. The attacker then downloads the benign model $M_{\bm{\theta}}$ and uses the data $D_{Adv}$ to fine-tune it to inject backdoor, resulting in $M_{\hat{\bm{\theta}}}\left( \hat{\bm{x}} \right)$ being assigned with a specific label $\hat{y}$, where $M_{\bm{\theta}}$ denotes the infected model. Subsequently, the attacker uploads the infected model to model markets, and unsuspected users may download and deploy it on their devices, such as an access control system. Once the malicious model is deployed, for instance, in an access control system, an unknown individual Alice can be erroneously recognized as the registered individual Candy, thereby enabling Alice to bypass the access control system and potentially leading to severe consequences.

\subsection{Attacker's Capabilities}
It is reasonable to assume that the attacker has access to the benign model, including its structure and parameters. In contrast to most of the existing related works~\cite{gu2019badnets, chen2017targeted, zhong2020backdoor, hammoud2021check}, we assume that the attacker can not get access to the benign model's training dataset or a dataset with identical distribution. In the majority of real-world scenarios, the training dataset consists of private data collected by the model owner, which may include sensitive information such as photos and addresses. Therefore, the model publisher will not release the corresponding training dataset, such data cannot be disclosed to the attacker. As a result, the attacker possesses only the minimum knowledge necessary for executing the proposed attack.

\section{Preliminaries} \label{sec:preliminary}
In this section, we introduce the main techniques adopted by this work, including Knowledge Distillation (KD) \cite{gou2021knowledge}, Gradient based Class Activation Mapping (Grad-CAM) \cite{selvaraju2017grad} and Elastic Weight Consolidation (EWC) \cite{kirkpatrick2017overcoming}.

\subsection{Knowledge Distillation}
Knowledge Distillation (KD) is an important technique for model compression, where a large-scale model acts as a teacher model and a compressed, lightweight model serves as a student model. The student model learns from the teacher model by matching the output distribution of the teacher model. This is achieved through the following loss function:
\begin{equation}
    L_{D}\left(\bm{y}, \hat{\bm{y}} \right) = -\sum_{k=1}^K y_{i}^{'} \cdot \mathrm{log}\left( \hat{y_{i}^{'}}\right), \nonumber
\end{equation}
where $\bm{y}$ and $\hat{\bm{y}}$ are the logits score output of the teacher and the student model respectively. $y_i^{'}$ and $\hat{y_i}^{'}$ are computed via
\begin{equation}
\label{distillation}
y_i^{'} = \frac{\mathrm{exp}\left( y_{i}/{T}\right)}{\sum_{k=1}^K \mathrm{exp}\left( y_{k}/{T}\right)} \; \mathrm{and} \; \hat{y_i^{'}} = \frac{\mathrm{exp}\left( \hat{y_{i}}/{T}\right)}{\sum_{k=1}^K \mathrm{exp}\left( \hat{y_{k}}/{T}\right)},
\end{equation}
where $T$ is the temperature parameter, $K$ is the number of classes, $y_{i}$ and $\hat{y_i^{'}}$ are the elements of tensors $\bm{y}$ and $\hat{\bm{y}}$, respectively. Compared with directly learning from the training set with hard labels, the KD manner provides more informative supervision signals from the teacher model which contains many knowledge learned by the teacher model and thus will benefit the training of the student model. The class distribution output reflects the resemblance of the input image to various image classes, providing valuable information to the student model, hence the student model can learn more effectively compared to learning from the training set with hard labels.
\subsection{Grad-CAM}
\begin{figure}[t]
\centering
	\begin{minipage}{0.3\linewidth}
		\vspace{3pt}
		\centerline{\includegraphics[width=\textwidth]{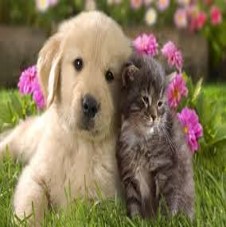}}
		\centerline{(a)}
	\end{minipage}
	\begin{minipage}{0.3\linewidth}
		\vspace{3pt}
		\centerline{\includegraphics[width=\textwidth]{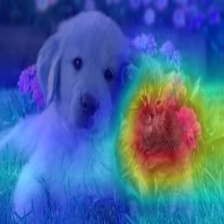}}
		\centerline{(b)}
	\end{minipage}
	\begin{minipage}{0.3\linewidth}
		\vspace{3pt}
		\centerline{\includegraphics[width=\textwidth]{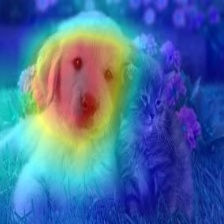}}
		\centerline{(c)}
	\end{minipage}
	\caption{Illustration of Grad-CAM. (a) is the input image of a Resnet50 model. (b) is the attention map corresponds to label `cat' and (c) is the attention map corresponds to label `dog'. Grad-CAM explains which pixels contribute most for model's prediction.}
	\label{fig:cam_demo}
 \end{figure}

Grad-CAM is a widely adopted model interpretation tool that provides visual insights into the pixels responsible for the classification of an input image into a specific class $c$, thereby explaining the model's output. To estimate the contribution of individual pixels, Grad-CAM first forwards the model and get the prediction score for each class. Subsequently, backpropagation is performed to compute the gradients of the score $y^{c}$ with respect to each convolutional feature map $A_k$ and the final feature map is obtained by 
\begin{equation}
\label{gradcam}
    ReLU\left( \sum_k \alpha_k^{c} A_k \right).
\end{equation}
More precisely, $\alpha_k^c = \frac{1}{Z}\sum_i\sum_j\frac{\partial y^c}{\partial A_{ij}^k}$ is the Global Average Pooling result of the gradient of $y^c$ with respect to $A_k$. In essence, Grad-CAM computes a weighted average of the convolutional feature maps $A_k$, where the importance of each feature is determined by its corresponding gradient. The demonstration of Grad-CAM is shown in Figure~\ref{fig:cam_demo}.
\subsection{Elastic Weight Consolidation}
Elastic Weight Consolidation (EWC) is a commonly adopted approach for continual learning~\cite{wang2024comprehensive}. In many real-world applications, a deep neural network model trained for task $A$ with dataset $D_A$ may be asked to deal with another related task $B$ with dataset $D_B$. However, directly fine-tuning the pre-trained model with dataset $D_B$ can result in catastrophic forgetting~\cite{goodfellow2013empirical} phenomenon, where the model's performance on task $A$ deteriorates significantly. EWC, based on a Bayesian framework for continual learning, leverages the Bayesian Theorem to calculate the posterior of the model's parameters, given by:
\begin{equation}
    \resizebox{\linewidth}{!}{$
    \displaystyle
    \begin{aligned}
    \mathrm{log}p\left(\theta \mid D \right) &= \mathrm{log}p\left( D_A, D_B \mid \theta\right) + \mathrm{log}p\left( \theta \right) - \mathrm{log}p\left( D_A, D_B\right) \\
    &= \mathrm{log}p\left( D_B \mid \theta \right) + \mathrm{log}p\left( \theta \mid D_A\right) - \mathrm{log}p\left( D_B\right). \label{ewc_bayes}
    \end{aligned}
    $}
\end{equation}
Equation~\eqref{ewc_bayes} indicates that, to maximize the posterior of $\theta$, one must maximize $\mathrm{log}p\left( D_B \mid \theta \right)$, which corresponds to the loss function for task $B$, while also maximizing the posterior given the dataset of task $A$, which is a regularization of parameters given by task $A$.

Directly maximizing the posterior $\mathrm{log}p\left( \theta \mid D_A\right)$ is intractable, therefore, Laplace Approximation \cite{mackay1992practical} technique is adopted to construct a surrogate objective. The idea of Laplace Approximation is to use a Gaussian distribution to approximate a complex distribution. The mean of the Gaussian distribution is the maximum of the approximated distribution $\theta_A^{*}$. We assume the model had converged on task $A$, and thus we can use the parameters of model trained with $D_A$ to approximate $\theta_A^{*}$. The co-variance matrix is the Hessian matrix of the negative log likelihood of $D_A$ at $\theta_A^{*}$:
\begin{equation}
\label{hessian}
    H_{ij} = -\left[ \frac{\partial^{2} \mathrm{log}p\left( \theta \mid D_A\right)}{\partial i \partial j}\mid_{\theta=\theta_A^{*}} \right].
\end{equation}
EWC keeps the diagonal elements $H_{ii}$ to estimate the importance of the parameters, and the loss function of EWC can be formulated as
\begin{equation}
\label{emc_loss}
    \mathcal{L} = \mathcal{L}_B\left( \theta \right) + \sum_{i}\frac{\lambda}{2}H_{ii}\left( \theta_{i} - \theta_{A, i}^{*} \right)^{2},
\end{equation}
where $\mathcal{L}_B$ is the loss for task $B$, $\lambda$ reflects the importance weight for the old task $A$.

\begin{figure*}
    \centering
    \includegraphics[width=\linewidth]{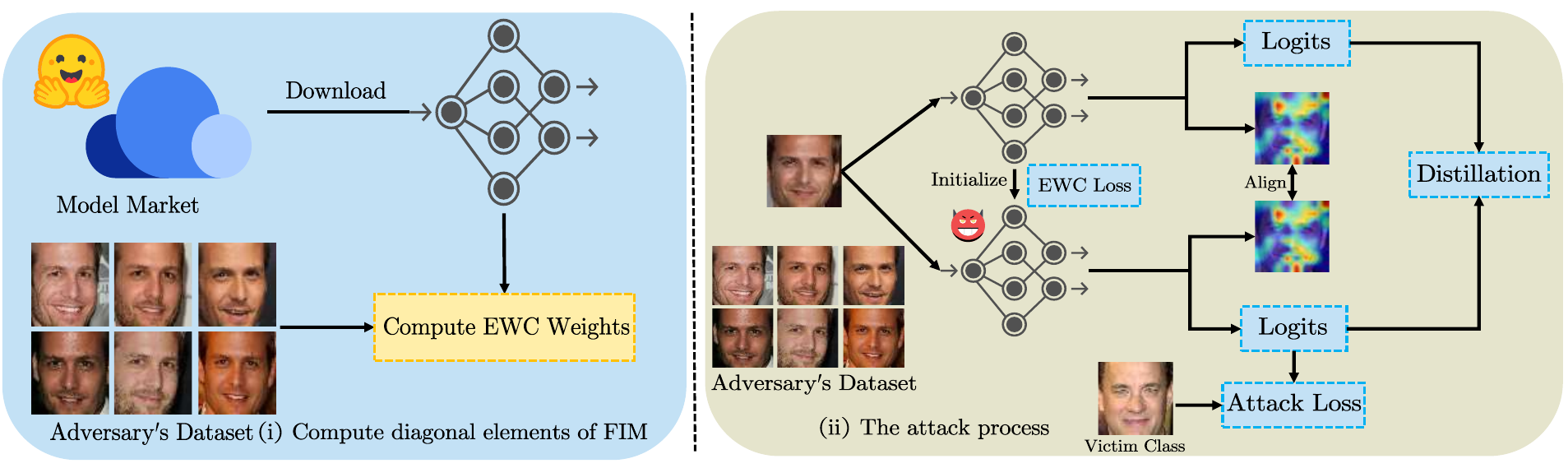}
    \caption{The overall process of the proposed attack. As shown in (i), the attacker first estimates the importance of the parameters of the pre-trained model via EWC method, and (ii) conducts the attack performance via the estimated EWC weights and distillation losses. Note that the attacker can only get access to $D_{Adv}$ without any training data of the pre-trained model.}
    \label{fig:process}
\end{figure*}

\section{The Proposed Attack} \label{sec:attack}
In this section, we firstly elucidate the technological challenges and then introduce design details associated with the proposed attack.

\subsection{Technological Challenges}
First of all, the naive method for conducting the proposed attack is directly training the benign model on the attacker's dataset $D_{Adv}=\left\{ \hat{\bm{x}}_1, \cdots, \hat{\bm{x}}_n \right \}$ with the label $\hat{y}$ desired by attacker. However, due to the lack of the original training dataset, such simplistic fine-tuning method can lead to a significant decrease in classifier performance on benign data due to the phenomenon of catastrophic forgetting \cite{goodfellow2013empirical}. 

Additionally, unlike existing backdoor attacks that rely on triggers to activate attacker-specified behaviors, the proposed attack does not require any trigger. Trigger-based backdoor attacks suffer from a lack of stealthiness, hence are prone to be detected and removed by current defense mechanisms~\cite{doan2020februus, zeng2021adversarial, wang2019neural,chen2022breaking, fu2023differential}. In contrast, the proposed trigger-free backdoor attack method operates without the need for any trigger, substantially enhancing the stealthiness of the attack. How to activate the backdoor implanted in the infected model without a specific trigger is another technical challenge we need to address. However, owing to its trigger-free nature, the proposed attack presents a greater challenge in achieving accurate triggering compared to existing backdoor attacks where the infected model can easily memorize the trigger pattern. Consequently, the attack success rate of the trigger-free backdoor attack may be relatively lower than that of other trigger-based backdoor attacks.

\subsection{Attack Design}
Figure~\ref{fig:process} illustrates the overview of the proposed attack. The attacker first downloads a pre-trained benign model from the model market, and uses $D_{Adv}$ to estimate the EWC weights of parameters of the benign model for preserving the knowledge learned by the model on benign samples. Afterwards, the attacker starts to inject backdoor into the benign model. To this end, the attacker should achieve two goals, i.e., misclassifying the malicious samples to the attacker-specified class while maintaining the consistency between infected model and benign model on the benign samples. To achieve the first goal, the attacker uses $D_{Adv}$ to compute the attack loss for injecting the backdoor into the infected model. For the second goal, the infected model is initialized via the benign model, and $D_{Adv}$ is fed to both the infected and benign models to compute the distillation losses to maintain the performance of the infected model on benign samples. Next, we introduce the proposed attack process in more details.

Because the attacker's goal is to fine-tune a benign model such that the generated infected model classifies the malicious input $\hat{\bm{x}}$ into the attacker-specific class with label $\hat{y}$, the trivial approach is to directly fine-tune the pre-trained model via the following loss function:
\begin{equation}
\label{attack_loss}
    \mathcal{L}\left( M_{\hat{\bm{\theta}}}\left( \hat{\bm{x}}\right), \hat{y}\right) = -\sum_{k=1}^K \mathrm{log}\left(\bm{y}_k\right) \cdot \hat{y} \cdot \mathbb{I}_{k=\hat{y}},
\end{equation}
where $\bm{y} = M_{\hat{\bm{\theta}}}\left( \hat{\bm{x}}\right)$ is the output of the infected model on malicious sample $\hat{\bm{x}}$ and $K$ is the number of classes. Though directly fine-tuning the infected model via Equation~\eqref{attack_loss} leads to high attack success rate, however, the fine-tuned model is prone to classifying all images, including benign ones, into the same class to achieve the attacker's goal. This is because only the weights connected to the output neurons of the label $\hat{y}$ are updated in the classification layer parameters during the fine-tuning process. Therefore, in the prediction process, the infected model has a preference for $\hat{y}$ in its classification result for a given input. Consequently, the model's performance decreases drastically on the benign input. To overcome this challenge, we adopt two types of regularizations to preserve the infected model's performance on benign samples, including distillation losses that align the infected and original model's outputs on benign samples, and EWC regularization that constrains the infected model's parameters not to move far away from the original parameters, hence maintaining the knowledge learned from the benign samples. 

Note that we assume the unavailability of the original training data, which further poses a challenge for the attacker in maintaining the infected model's performance on benign samples. To address this issue, we leverage the malicious samples $D_{Adv}$ for knowledge distillation. More concretely, we maintain a copy of the original model $M_{\bm{\theta}}$ as the teacher model, while the infected model $M_{\hat{\bm{\theta}}}$ serves as the student model. For $\hat{\bm{x}} \in D_{Adv}$, the distillation loss is formulated as:
\begin{equation}
    L_{D}\left(\bm{y}, \hat{\bm{y}} \right) = -\sum_{k=1}^K \bm{y}_{i}^{'} \cdot \mathrm{log}\left( \hat{\bm{y}}_{i}^{'}\right),
\end{equation}
where $\bm{y}=M_{\bm{\theta}}\left( \hat{\bm{x}}\right)$ and $\hat{\bm{y}}=M_{\hat{\bm{\theta}}}\left( \hat{\bm{x}}\right)$ are the prediction vectors (composed of logits scores) of the benign model and the infected model respectively. Besides, $\bm{y}_i^{'}$ and $\hat{\bm{y}}_i^{'}$ are calculated via Equation~\eqref{distillation}. Incorporating the knowledge distillation loss in the proposed attack serves to constrain the fine-tuned model to behave consistently with the original model on benign samples. Therefore, the infected model can inherit the learned knowledge from the benign model and maintain performance on benign data.

While the knowledge distillation loss can alleviate the performance degradation of the infected model, it fails to take the image details into account. Pixels with significant influence on the model's prediction constitute the attention region of the network, and the attention regions encode model's representation more precisely. Instead of learning which classes are resembled in the input data, attention maps explain why the model outputs the prediction. Grad-CAM provides a method to visualize which region contributing most to the model's predictions and thus provides an explanation for the model's output. Given the image $i$, the model $M_{\bm{\theta}}$, and class $c$ with the highest prediction score, the Grad-CAM attention map vector is formulated as
\begin{equation}
    Q_{\bm{\theta}}^{i,c} = \bm{\mathrm{vector}}\left( \bm{\mathrm{Grad\text{-}CAM}}\left( i, M_{\bm{\theta}}\left(\hat{\bm{x}} \right), c \right) \right),
\end{equation}
where $\bm{\mathrm{vector}}\left(\cdot \right)$ indicates the vectorization operation and $\bm{\mathrm{Grad\text{-}CAM}}\left( i, M_{\bm{\theta}}\left(\hat{\bm{x}} \right), c \right)$ is calculated via Equation~\eqref{gradcam}. 
In our scheme, in addition to utilizing knowledge distillation to maintain the performance of the infected model on benign inputs, we also consider maintaining the consistency of the infection model against the interpretation results of the input data. The attention distillation loss is defined as
\begin{equation}
    L_{AD} = \sum_{j=1}^{l} \Vert\frac{Q_{\bm{\theta}, j}^{i, c}}{\Vert Q_{\bm{\theta}}^{i, c}\Vert_2} - \frac{Q_{\hat{\bm{\theta}}, j}^{i, c}}{\Vert Q_{\hat{\bm{\theta}}}^{i, c}\Vert_2}\Vert_1,
\end{equation}
where $l$ is the length of the attention mapping vector. By incorporating attention mapping regularization, the infected model is compelled not only to output a similar class probability distribution as the original model but also to align their interpretations. This results in a more comprehensive knowledge preservation regularization.

\begin{algorithm}[t]
    \caption{The Attack Process}
    \label{alg:attack}
    \SetKwInOut{Input}{Input}
    \SetKwInOut{Output}{Output} 
    \Input{The pre-trained model $M_{\bm{\theta}}$, the adversary's dataset $D_{Adv}=\left\{ \hat{\bm{x}}_1, \cdots, \hat{\bm{x}}_N \right\}$, the label of the target class $t$, and the learning rate $\eta$ for attacking process.}
    \Output{The backdoored model $M_{\hat{\bm{\theta}}}$}
    \BlankLine
    $ \bm{f} \leftarrow \bm{0}$ \Comment{Estimate the diagonal elements of FIM} \\
    \For{$\hat{\bm{x}}$ in $D_{Adv}$}{
        $\bm{y}=M_{\bm{{\theta}}}\left( \hat{\bm{x}}\right)$ \\
        $y^{*} = \mathbf{argmax}\left( \bm{y}\right)$  \Comment{Approximate $\bm{\mathrm{softmax}}\left( \hat{\bm{y}}\right)$}\\
        $\mathcal{L} = \mathbf{cross\_entropy}\left( \bm{y}, y^{*}\right)$ \\
        $\bm{g} = \frac{\partial \mathcal{L}}{\partial \bm{\theta}}$ \\
        \For{$i$ in $\left| D_{Adv}\right|$}{
            $\bm{f}_{i} = \bm{f}_i + \bm{g}_i^2$ \Comment{Estimate FIM via Eq.(\ref{f_ii})}
        }
    }
    $\bm{f} = \frac{\bm{f}}{\left| D_{Adv}\right|}$\\
    $M_{\hat{\bm{\theta}}} \leftarrow M_{\bm{\theta}}$ \Comment{Initialize the victim model}\\
    \For{epoch in num\_epochs}{
        $L \leftarrow 0$ \Comment{Start the attack loop} \\
        \For{$\hat{\bm{x}}$ in $D_{Adv}$}{
            $\bm{y}=M_{\bm{\theta}}\left( \hat{\bm{x}}\right)$ \\
            $y^{*} = \mathbf{argmax}\left( \bm{y}\right)$\\
            $\hat{\bm{y}} = M_{\hat{\bm{\theta}}}\left( \hat{\bm{x}}\right)$ \\
            $\mathcal{L} = \mathbf{cross\_entropy}\left( \hat{\bm{y}}, t\right)$ \Comment{Attack loss}\\
            $L_D=-\sum_{k=1}^K \bm{y}_{i} \cdot \mathrm{log}\left( \hat{\bm{y}}_{i}\right)$ \Comment{Distillation loss}\\
            // Grad-CAM attention distillation loss \\
            $\bm{Q}_{\bm{\theta}}^{\hat{\bm{x}}, y^{*}} = \mathbf{vector}\left(\mathbf{Grad\text{-}CAM}\left(\hat{\bm{x}}, M_{\bm{\theta}}, y^{*} \right) \right)$ \\
            $\bm{Q}_{\hat{\bm{\theta}}}^{\hat{\bm{x}}, y^{*}} = \mathbf{vector}\left(\mathbf{Grad\text{-}CAM}\left(\hat{\bm{x}}, M_{\hat{\bm{\theta}}}, y^{*} \right) \right)$ \\
            $L_{AD} = \sum_{j=1} \Vert\frac{\bm{Q}_{\bm{\theta}, j}^{\hat{\bm{x}}, y^{*}}}{\Vert \bm{Q}_{\bm{\theta}}^{\hat{\bm{x}}, y^{*}}\Vert_2} - \frac{\bm{Q}_{\hat{\bm{\theta}}, j}^{\hat{\bm{x}}, y^{*}}}{\Vert \bm{Q}_{\hat{\bm{\theta}}}^{\hat{\bm{x}}, y^{*}}\Vert_2}\Vert_1$ \\
            $L = L + \mathcal{L} + \alpha L_D + \beta L_{AD}$
        }
        $L_{EWC} \leftarrow 0$ \Comment{Loss for EWC regularization}\\
        \For{$i$ in $\left| \bm{\theta}\right|$}{
            $L_{EWC} = L_{EWC} + \frac{\lambda}{2}\bm{f}_i\left( \hat{\bm{\theta}}_i-\bm{\theta}_i\right)^2$
        }
        $L = L + L_{EWC}$ \\
        $\hat{\bm{\theta}}=\hat{\bm{\theta}}-\eta\frac{\partial L}{\partial \hat{\bm{\theta}}}$ \Comment{Backpropagation and update}
    }
    \textbf{return} $M_{\hat{\bm{\theta}}}$
\end{algorithm}

Though the distillation loss and attention distillation loss can mitigate catastrophic forgetting to some extent, a more straightforward approach is to directly constrain the important parameters of the fine-tuned model from changing significantly. However, not all parameters have equal importance, as some are crucial to the attacker's goal while others are not. To address this, we weight the importance of parameters and impose stronger regularization to parameters with greater importance. This ensures that important parameters remain unchanged, while the remaining parameters can be utilized for the attacker's task. To quantify parameter importance, we leverage the EWC algorithm. As shown in Equation~\eqref{emc_loss}, EWC employs Laplace approximation to estimate the posterior and need to compute the Hessian matrix via Equation~\eqref{hessian}. However, directly computing Hessian matrix via Equation~\eqref{hessian} requires high computational cost due to the great dimension of $\mathbb{R}^{\mid\theta\mid}$. To overcome this issue, the Hessian is approximated by Fisher Information Matrix (FIM):
\begin{equation}
\label{fim}
    F = \mathbb{E} \left[ \nabla_{\theta} \mathrm{log} \; p\left( D_A \mid \theta \right) \nabla_{\theta} \mathrm{log} \; p\left( D_A \mid \theta \right)^{T} \right] \Big|_{\theta=\theta_{A}^{*}} \approx H. \nonumber
\end{equation}
By leveraging Equation~\eqref{fim}, we can use the Monte Carlo method to estimate the expectation. Specifically, for the proposed attack, the adversary only possesses the dataset $D_{Adv}$ used for injecting the backdoor into a pre-trained model and does not have access to any data from the training set of the victim model. Therefore, the attacker can only use $\hat{\bm{x}} \in D_{Adv}$ to estimate the importance of parameters:
\begin{equation}
\label{f_ii}
    F_{ii} = \frac{1}{N} \sum_{\hat{\bm{x}} \in D_{Adv}} \left( \frac{\partial L\left(\bm{y}, y^{*}\right)}{\partial \bm{\theta}_i}\right)^{2},
\end{equation}
where L is the loss function (cross entropy) and $\hat{\bm{y}}$ is the output of $M_{\bm{\theta}}\left( \hat{\bm{x}}\right)$, $y^{*} \sim \mathrm{Multimonial}\left(\mathrm{softmax}\left( \bm{y}\right)\right)$ is the label sampled from the model's prediction. Leveraging Equation~\eqref{f_ii}, the importance weights of the parameters can be efficiently computed through backpropagation. The EWC regularization term then can be formulated as:
\begin{equation}
    L_{EWC} = \sum_{i} \frac{\lambda}{2} F_{ii} \left( \hat{\bm{\theta}}_i - \bm{\theta}_i \right)^2, \nonumber
\end{equation}
where $\bm{\theta}_i$ is the $i$-th component of parameters of the benign pre-trained model and $\lambda$ is the hyperparameter.

In summary, our overall optimization objective can be formulated as follows:
\begin{equation}
\resizebox{0.91\linewidth}{!}{$
    \displaystyle
    \begin{aligned}
        \mathop{\min}_{\hat{\bm{\theta}}} L &= \sum_{\hat{\bm{x}} \in D_{Adv}}\mathcal{L}\left( M_{\hat{\bm{\theta}}}\left( \hat{\bm{x}}\right), \hat{y}\right) + \alpha \sum_{\hat{\bm{x}} \in D_{Adv}} L_D\left( M_{\bm{\theta}}\left( \hat{\bm{x}}\right),  M_{\hat{\bm{\theta}}}\left( \hat{\bm{x}}\right)\right) \\
        &+ \beta \sum_{\hat{\bm{x}} \in D_{Adv}}L_{AD}\left( M_{\bm{\theta}}\left( \hat{\bm{x}}\right), M_{\hat{\bm{\theta}}}\left( \hat{\bm{x}}\right)\right) + L_{EWC}. \nonumber
    \end{aligned}
$}
\end{equation}
The first term is used to classify the malicious data to the specified label $\hat{y}$, the second term ensures the consistency of the infected model and the benign model on the benign input prediction, the third ensures the correctness of the infection model for the interpretation results of the input, and the fourth makes the infected model more efficient to complete the training by quantifying the importance of parameters. The entire attack process is summarized in Algorithm~\ref{alg:attack}. Lines 1 to 11 are used to quantify the parameter importance using EWC, and lines 12 to 33 calculate the proposed loss functions, $\mathcal{L}$, $L_D$, and $L_{AD}$, respectively to form the total loss $L$ for training the infected model $M_{\hat{\bm{\theta}}}$.

\section{Experiments} \label{sec:evaluation}
In this section, we conducted comprehensive experiments to evaluate the performance of the proposed backdoor attack. All the experimetns are conducted on a server with an Nvidia RTX 3090 GPU, an Intel(R) Xeon(R) CPU E5-2699Cv4 @ 2.20GHz, 64G RAM and Ubuntu 20.04 system. We adopted Python 3.8.18 as our programming language and Pytorch 1.13 as the training backend.
\subsection{Experimental Setup}
\noindent \textbf{Datasets \& Models:} The experiments are counducted on several benchmark datasets (i.e., Cifar100~\cite{krizhevsky2009learning}, FaceScrub~\cite{ng2014data}, MNIST~\cite{lecun1998gradient}) to evaluate the effectiveness of the proposed attack. We briefly introduce the datasets we adopted for our experiments as follows:
\begin{itemize}
    \item \textbf{Cifar100:} Cifar100 consists of 60,000 $\times$ 32 color images, categorized into 100 classes, e.g., airplane, automobile and so on. Cifar100 is uniformly distributed, and each class consists of 60 images, 50 as the training data and 10 as the test data. And we adopted ResNet32 to classify the images in Cifar100.
    \item \textbf{FaceScrub:} FaceScrub consists of 48,579 images from 530 individuals. FaceScrub is a commonly adopted dataset for facial recognition task. We used ResNet34 to classify the photos in FaceScrub.
    \item \textbf{MNIST:} MNIST is a database of handwritten digits from 0-9 that is commonly used for training various image processing systems. The MNIST dataset contains 60,000 training images and 10,000 testing images. We used LeNet to classify the images in MNIST.
\end{itemize}

Before attacking the neural networks, we first train neural networks on the adopted datasets. We kept part of the classes of each dataset for pre-training, and recorded the corresponding accuracy on benign samples. The information of datasets, the adopted neural network models, and the corresponding accuracy on benign samples are summarized in Table~\ref{tab:dataset}.\\
\begin{table}[htb]
\begin{center}
\caption{The datasets and models for evaluation.}
\label{tab:dataset}
\begin{tabular}{| c | c | c | c | c |}
\hline
Dataset & Model & Classes& Accuracy\\
\hline
Cifar100 & ResNet32 & 80 & 0.631\\
\hline
FaceScrub & ResNet34 & 20 & 0.822\\ 
\hline
MNIST & LeNet & 9 & 0.987\\
\hline 
\end{tabular}
\end{center}
\end{table}

\noindent \textbf{Metrics:} We mainly evaluate the proposed backdoor attack using the following metrics:
\begin{enumerate}
    \item \textit{Attack Success Rate}: Attack success rate (ASR) quantifies the percentage of malicious samples which can successfully misclassified as the target label $t$. The ASR metric is defined via:
    \begin{equation}
        \mathrm{ASR} = \frac{\#\;\mathrm{successful\;misclassification}}{\#\;\mathrm{number\;of\;malicious\;inputs}}. \nonumber
    \end{equation}
    \item \textit{Accuracy Gap}: Accuracy Gap indicates the accuracy difference between the backdoored and benign models on benign samples, which can be formulated as:
    \begin{equation}
        \Delta \mathrm{Accuracy} = \sum_{\bm{x} \in \chi_{test}} \frac{\mathbb{I}_{M_{\bm{\theta}}\left(\bm{x} \right) \neq M_{\hat{\bm{\theta}}}\left( \bm{x}\right)}}{\left| \chi_{test}\right|},\nonumber
    \end{equation}
    where $M_{\bm{\theta}}$ and $M_{\hat{\bm{\theta}}}$ denotes the benign neural network and the backdoored neural network respectively, and $\chi_{test}$ indicates the test dataset of benign samples.
    \item \textit{Time Cost}: This metric aims to measure the time consumption to implement the attack. We demonstrate the time consumption for different dataset to validate the effectiveness of the proposed attack.
\end{enumerate}

\subsection{Basic Evaluation of the Attack}
This section we evaluate the effectiveness of the proposed attack, including the attack success rate (ASR), accuracy gap and time cost. We assume that the attacker can only get access to the malicious data used for backdoor injection. We evaluated the proposed attack on Cifar100, FaceScrub and MNIST dataset. For each dataset, we first train a neural network to classify the images within the dataset. For Cifar100 dataset, we kept the first 80 classes for pre-training, and we trained a ResNet32~\cite{he2016deep} network via Adam optimizer with learning rate of 1e-3. For FaceScrub dataset, we kept first 20 individual with most amount of pictures for pre-training and trained ResNet34~\cite{he2016deep} network to classify the pictures with Adam optimizer of learning rate 1e-3. And for MNIST dataset, we kept first 9 classes for pre-training, and a LeNet~\cite{lecun1998gradient} was trained with Adam optimizer of learning rate 1e-3.

\begin{figure}[htbp]
    \centering
    \includegraphics[width=\linewidth]{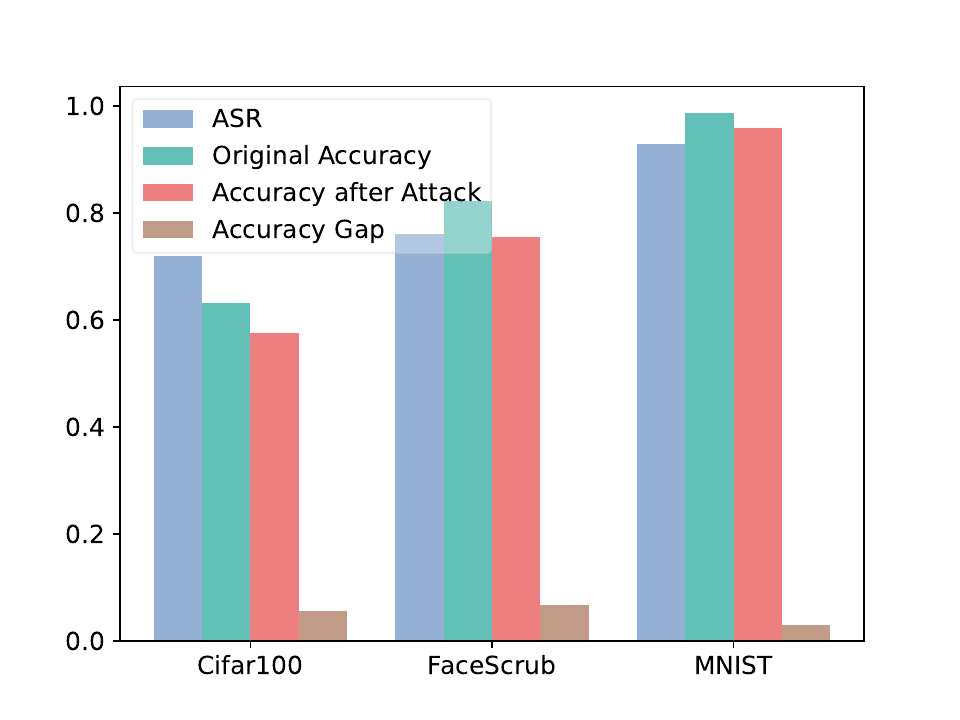}
    \caption{Basic evaluation of the proposed attack. }
    \label{fig:main_result}
\end{figure}

For attacking the pre-trained neural networks, we set the attack learning rate to be 5e-4 and $\alpha=4.0$, $\beta=10.0$, $\lambda=10000$ for Cifar100 dataset. For FaceScrub dataset, we set the attack learning rate to be 5e-4, and $\alpha=1.5$, $\beta=10.0$, $\lambda=50000$. For MNIST dataset, we the attacking learning rate to be 5e-4, and $\alpha=2.0$, $\beta=10.0$, $\lambda=10000$. For the malicious samples, we splitted them into two parts without intersection, one contains 80\% samples for backdoor injection, and the other one contains 20\% samples for ASR evaluation. The experimental results are shown in Figure~\ref{fig:main_result}.

\begin{table}[t]
\begin{center}
\caption{The time cost of the proposed attack.}
\label{tab:time_cost}
\begin{tabular}{| c | c | c | c | c |}
\hline
Dataset & Cifar100 & FaceScrub & MNIST \\
\hline
Model & ResNet32 & ResNet34 & LeNet \\
\hline
Time Cost (s) & 136.7 & 84.6 & 113.1 \\ 
\hline 
\end{tabular}
\end{center}
\end{table}

Because of the trigger-free nature of the proposed attack, it is more challenging for the model to memorize the malicious samples and activate the backdoor behavior compared to other backdoor attacks with triggers. In those attacks, the model can easily memorize the trigger. Therefore, we do not expect a very high ASR. As long as the ASR is comparable to the victim model's accuracy on the benign dataset, the attack will be considered complete. As shown in Figure~\ref{fig:main_result}, the proposed attack achieved a relatively high ASR while maintaining the performance of the victim model on benign samples. Additionally, the \textit{time costs} of the proposed attack are listed in Table~\ref{tab:time_cost}. The results indicate that the time cost of the proposed attack is quite low, further validating its effectiveness. The experimental results reveal a severe security threat to widely applied deep neural network models. The attacker can secretly inject a backdoor into a pre-trained neural network without access to the training data or training process of the victim model, while maintaining performance on the benign data. Besides, the trigger-free nature of the proposed model makes it more difficult to be detected by backdoor defense methods.

\subsection{Impacts of Hyper-parameters}
In this section, we analyze the impacts of hyperparameters on the proposed attack. Specifically, we examine the effects of the learning rate for the attack $\eta$, the temperature parameter for distillation $T$, the weight for distillation loss $\alpha$, and the weight for EWC loss $\lambda$. 

\begin{figure*}[htbp]
	\begin{minipage}{0.34\linewidth}
		\vspace{3pt}
		\centerline{\includegraphics[width=\textwidth]{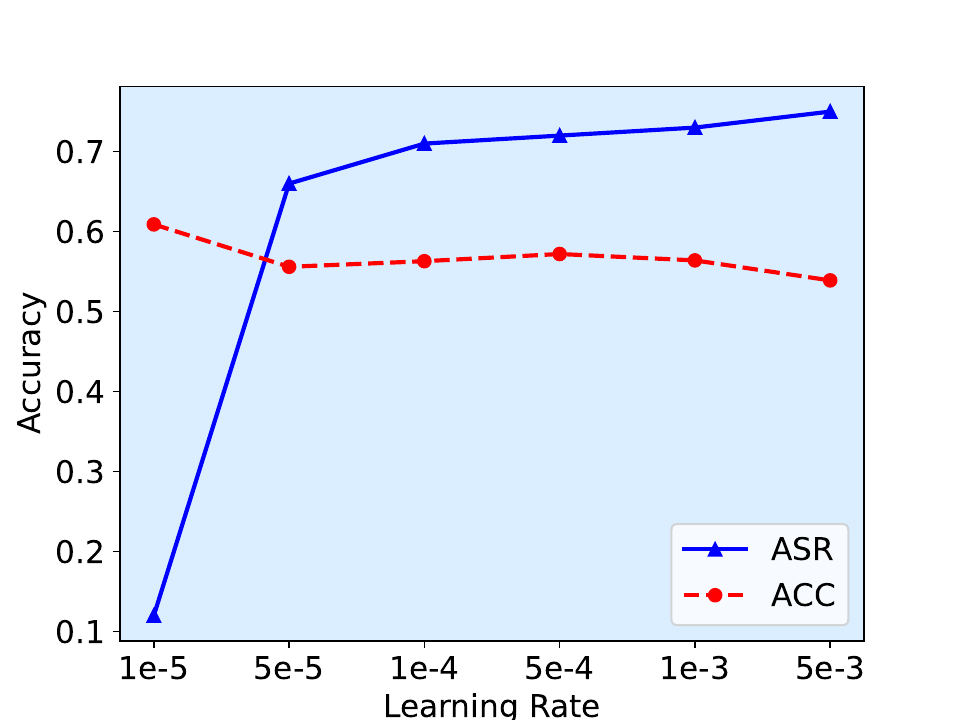}}
		\centerline{(a)}
	\end{minipage}
	\begin{minipage}{0.34\linewidth}
		\vspace{3pt}
		\centerline{\includegraphics[width=\textwidth]{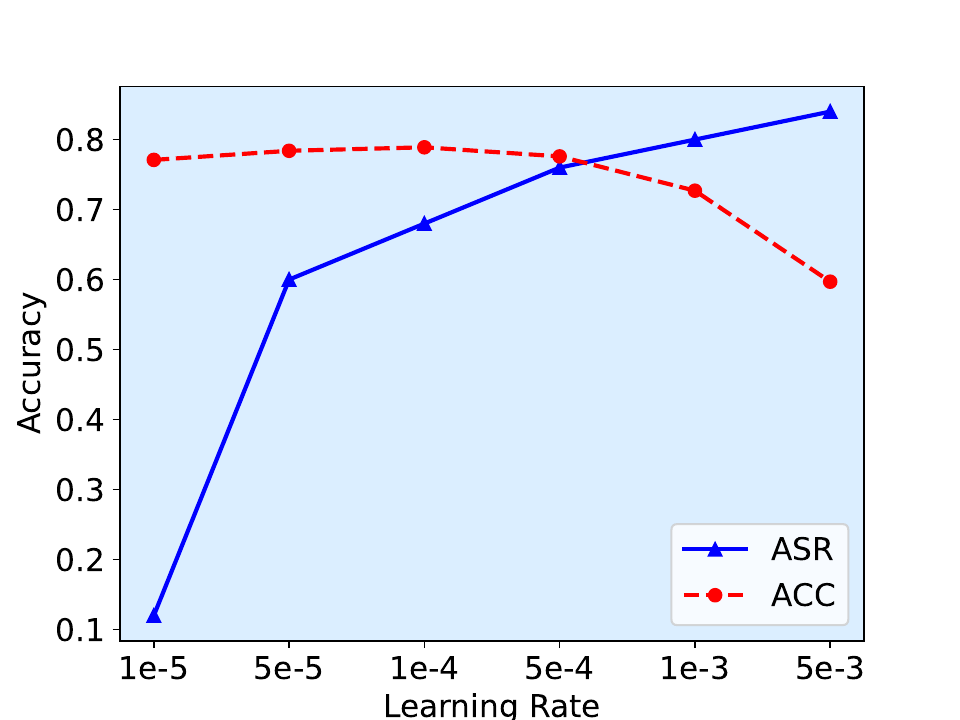}}
		\centerline{(b)}
	\end{minipage}
	\begin{minipage}{0.34\linewidth}
		\vspace{3pt}
		\centerline{\includegraphics[width=\textwidth]{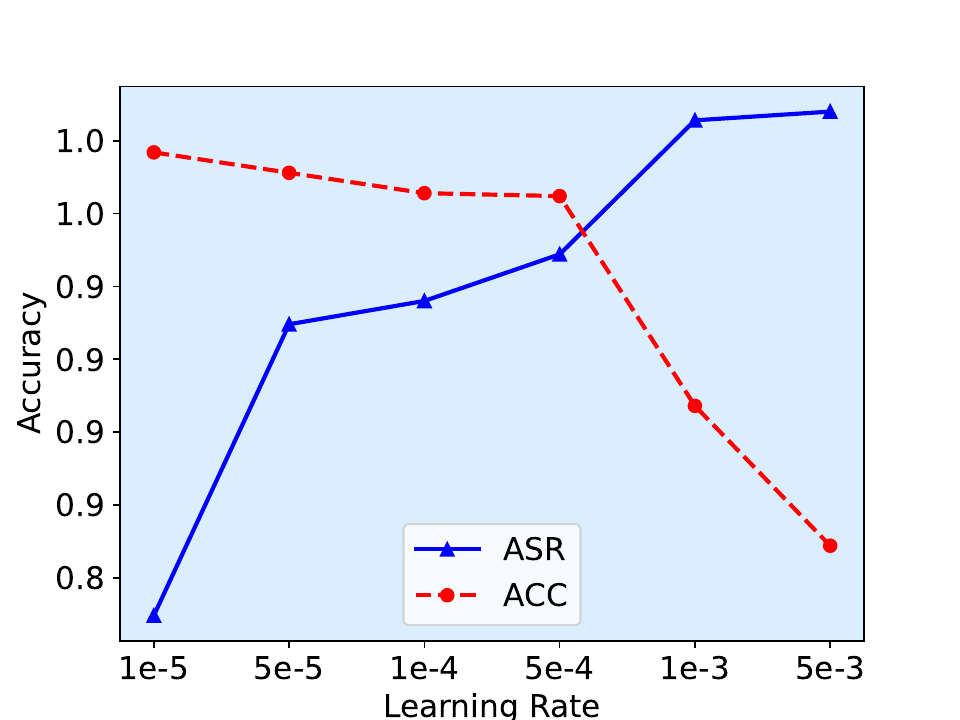}}
		\centerline{(c)}
	\end{minipage}
	\caption{Impact of learning rate for the proposed attack. (a)-(c) demonstrated how the ASR and accuracy of the infected model change with attack learning rate on various datasets. (a) demonstrates the performance change on Cifar100 dataset, where we set $\alpha=4.0$, $\beta=10.0$ and $\lambda=10000$. (b) demonstrates the performance change on FaceScrub dataset, where we set $\alpha=2.0$, $\beta=10.0$ and $\lambda=50000$. (c) depicts the performance change on MNIST dataset, where we set $\alpha=1.5$, $\beta=10.0$ and $\lambda=10000$.}
	\label{fig:lr}
\end{figure*}

First, we analyze the impacts of the learning rate for the attack, $\eta$. We evaluate the ASR and accuracy of the infected model for different attack learning rates. As depicted in Figure~\ref{fig:lr}, the attack learning rate has a significant effect on the attack performance. Generally, as the learning rate increases, the attack success rate also increases, while the accuracy of the infected model decreases. Within the range of learning rates from 1e-4 to 1e-3, the proposed attack demonstrates relatively stable and high performance. However, when the learning rate is too large (e.g., 5e-3), although the attack success rate can be very high, the accuracy of the infected model decreases significantly. Conversely, if the learning rate is too small (e.g., 1e-5), the infected model becomes trapped in the local minimum of the pre-trained model, resulting in a low attack success rate while maintaining a relatively high accuracy of the infected model.

\begin{figure*}[htbp]
	\begin{minipage}{0.34\linewidth}
		\vspace{3pt}
		\centerline{\includegraphics[width=\textwidth]{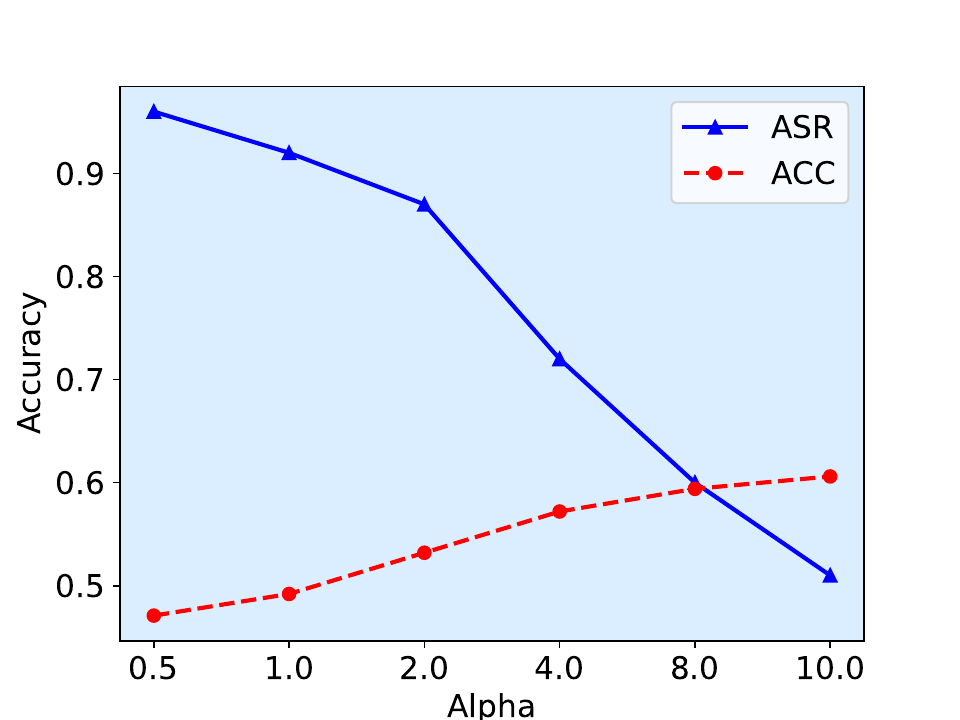}}
		\centerline{(a)}
	\end{minipage}
	\begin{minipage}{0.34\linewidth}
		\vspace{3pt}
		\centerline{\includegraphics[width=\textwidth]{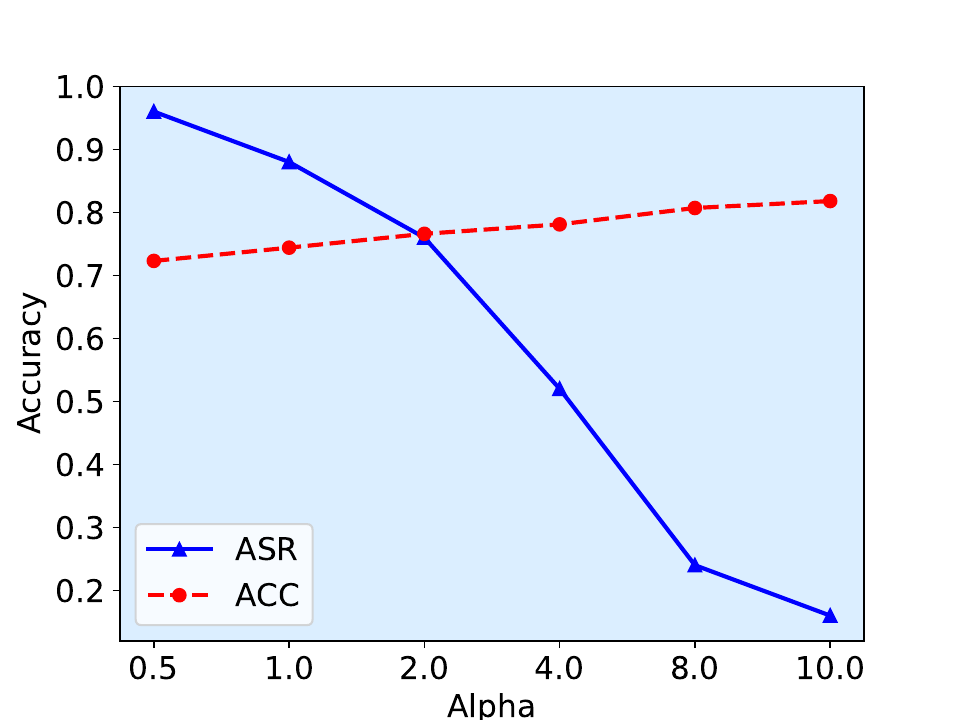}}
		\centerline{(b)}
	\end{minipage}
	\begin{minipage}{0.34\linewidth}
		\vspace{3pt}
		\centerline{\includegraphics[width=\textwidth]{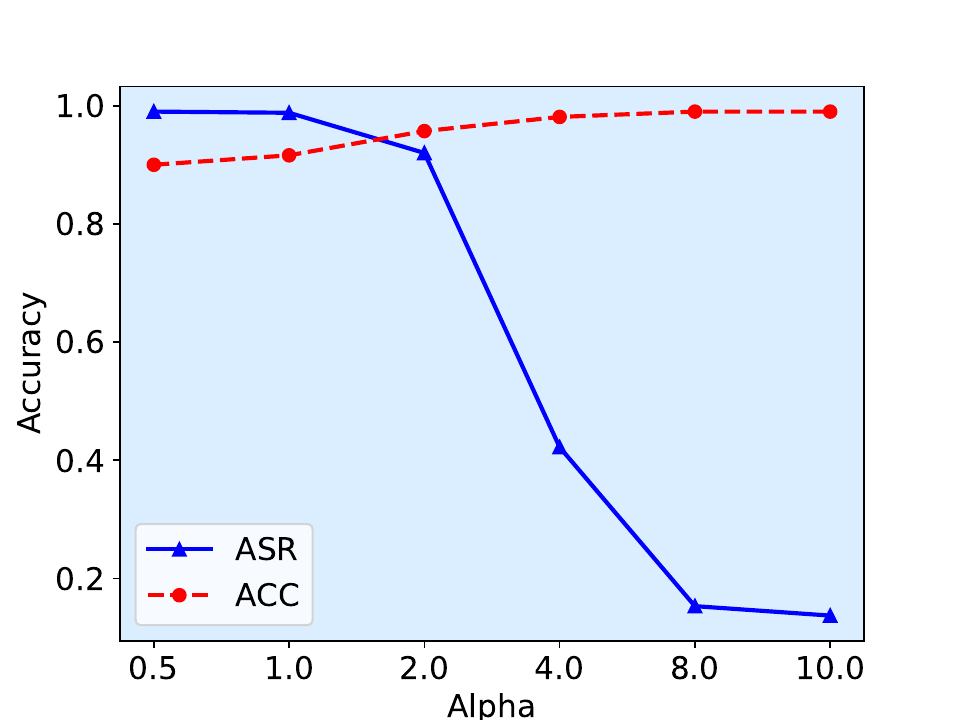}}
		\centerline{(c)}
	\end{minipage}
	\caption{Impact of hyper-parameter $\alpha$ for the proposed attack. (a)-(c) demonstrated how the ASR and accuracy of the infected model change with attack learning rate on various datasets. (a) demonstrates the performance change on Cifar100 dataset, where we set $\eta=0.0005$, $\beta=10.0$ and $\lambda=10000$. (b) demonstrates the performance change on FaceScrub dataset, where we set $\eta=0.0005$, $\beta=10.0$ and $\lambda=50000$. (c) depicts the performance change on MNIST dataset, where we set $\eta=0.0005$, $\beta=10.0$ and $\lambda=10000$.}
	\label{fig:alpha}
\end{figure*}

Next we analyzed the impacts of the weight for distillation loss $\alpha$. The curves of ASR and accuracy of infected model changes with alpha are depicted in Figure~\ref{fig:alpha}. We can see that the hyper-parameter $\alpha$ imposes important influences on the performance of the proposed attack. Specifically, with $\alpha$ increasing, the infected model will be imposed a stronger regularization that predict consistently with the benign model, hence keeps better performance on benign samples. However, as shown in Figure~\ref{fig:alpha}, imposing too strong distillation regularization will significantly decrease the attack success rate, that is partly because the attack dataset $D_{Adv}$ simultaneously adopted for both distillation loss and attack loss, and thus too strong distillation loss will undermine the attack loss.

\begin{figure*}[t]
	\begin{minipage}{0.34\linewidth}
		\vspace{3pt}
		\centerline{\includegraphics[width=\textwidth]{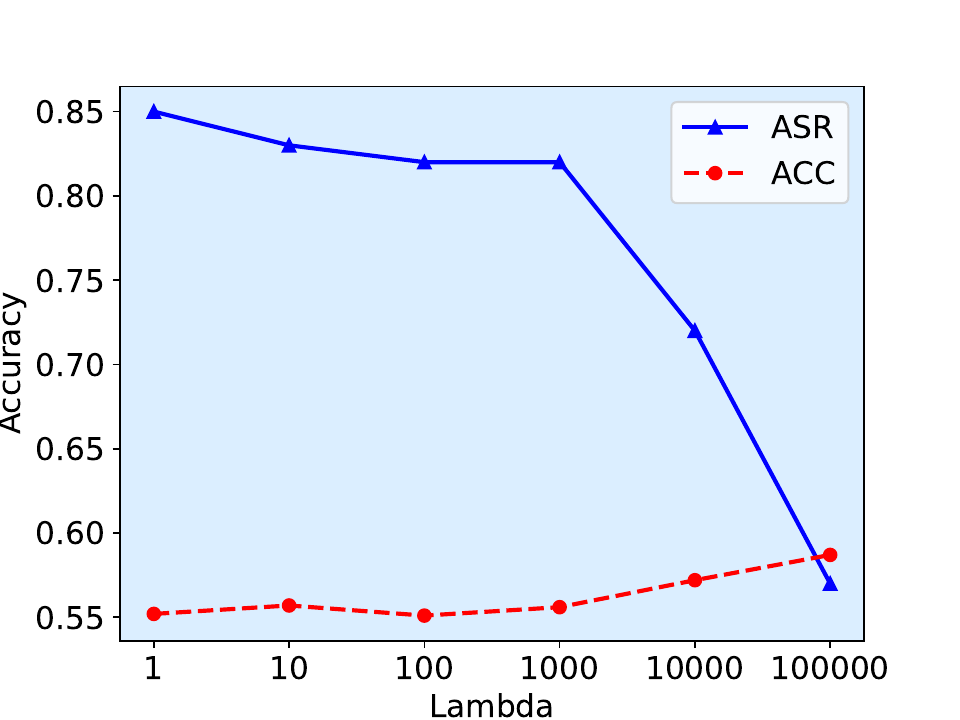}}
		\centerline{(a)}
	\end{minipage}
	\begin{minipage}{0.34\linewidth}
		\vspace{3pt}
		\centerline{\includegraphics[width=\textwidth]{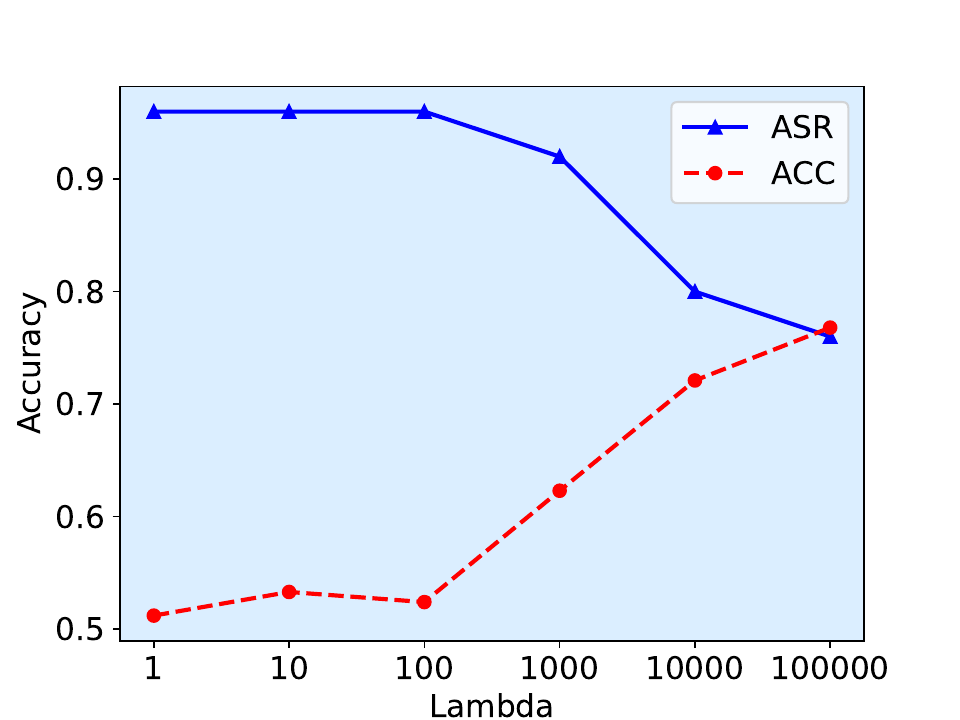}}
		\centerline{(b)}
	\end{minipage}
	\begin{minipage}{0.34\linewidth}
		\vspace{3pt}
		\centerline{\includegraphics[width=\textwidth]{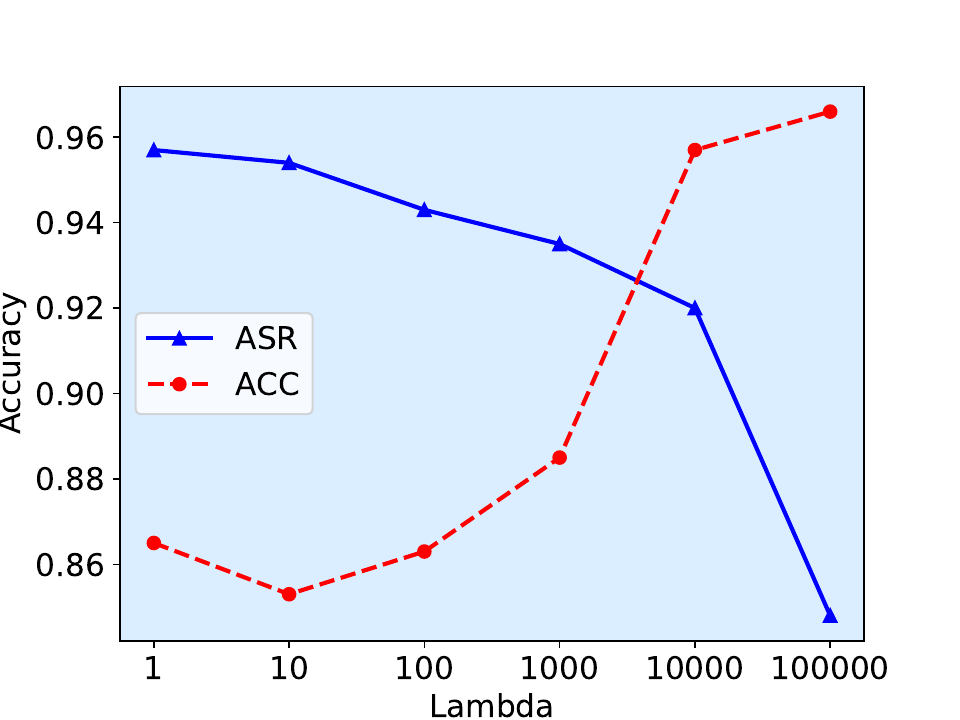}}
		\centerline{(c)}
	\end{minipage}
	\caption{Impact of hyper-parameter $\lambda$ for the proposed attack. (a)-(c) demonstrated how the ASR and accuracy of the infected model change with $\lambda$ on various datasets. (a) demonstrates the performance change on Cifar100 dataset, where we set $\alpha=4.0$, $\beta=10.0$ and $\eta=0.0005$. (b) demonstrates the performance change on FaceScrub dataset, where we set $\alpha=2.0$, $\beta=10.0$ and $\eta=0.0005$. (c) depicts the performance change on MNIST dataset, where we set $\alpha=1.5$, $\beta=10.0$ and $\eta=0.0005$.}
	\label{fig:lambda}
\end{figure*}

Similarly, the impacts of the weight for EWC loss, $\lambda$, are demonstrated in Figure~\ref{fig:lambda}. In contrast to $\alpha$, the performance of the proposed attack is not highly sensitive to variations in $\lambda$. However, when $\lambda$ is set to a large value (e.g., 100000), the attack success rate decreases while the accuracy of the infected model on benign samples remains high. Conversely, if $\lambda$ is small (e.g., 1, 10), the EWC regularization loses its effectiveness, resulting in a significant decrease in the accuracy of the infected model on benign samples.
\vspace{-1em}

\subsection{Backdoor Attack with Auxiliary Dataset}
So far, we have made the assumption that the attacker only has access to the malicious samples used for backdoor injection. While this assumption is realistic because it is typically impossible for the attacker to obtain the original training samples, the limited availability of such samples can restrict the attack performance. However, in many scenarios, the attacker can leverage publicly available datasets that have a similar distribution to the training set, thus assisting the attack. For instance, let's consider a scenario where the attacker aims to inject a backdoor into a facial recognition model. Although the attacker may not know which specific photos were used for model training, they can download facial datasets from the Internet to support the attack. By incorporating this auxiliary dataset, the attacker can expand the data used for the attack, thereby enhancing the attack performance. Concretely, to improve the performance of the proposed attack, we can use some auxiliary dataset for distillation loss to maintain the consistency of the infected model with the original one. Denoting the auxiliary dataset as $D_{Aux}$ and the new loss function can be formulated as: 
\begin{equation}
\resizebox{0.91\linewidth}{!}{$
    \displaystyle
    \begin{aligned}
        \mathop{\min}_{\hat{\bm{\theta}}} L &= \sum_{\hat{\bm{x}} \in D_{Adv}}\mathcal{L}\left( M_{\hat{\bm{\theta}}}\left( \hat{\bm{x}}\right), \hat{y}\right) + \alpha \sum_{\tilde{\bm{x}} \in D_{Aux}} L_D\left( M_{\bm{\theta}}\left( \tilde{\bm{x}}\right),  M_{\hat{\bm{\theta}}}\left( \tilde{\bm{x}}\right)\right) \\
        &+ \beta \sum_{\tilde{\bm{x}} \in D_{Aux}}L_{AD}\left( M_{\bm{\theta}}\left( \tilde{\bm{x}}\right), M_{\hat{\bm{\theta}}}\left( \tilde{\bm{x}}\right)\right) + L_{EWC}. \nonumber
    \end{aligned}
$}
\end{equation}
Different from the setting that the attacker can only get access to $D_{Adv}$, in this setting we leverage $D_{Adv}$ for attacking and $D_{Aux}$ for distillation losses.

\begin{figure*}[htbp]
	\begin{minipage}{0.34\linewidth}
		\vspace{3pt}
		\centerline{\includegraphics[width=\textwidth]{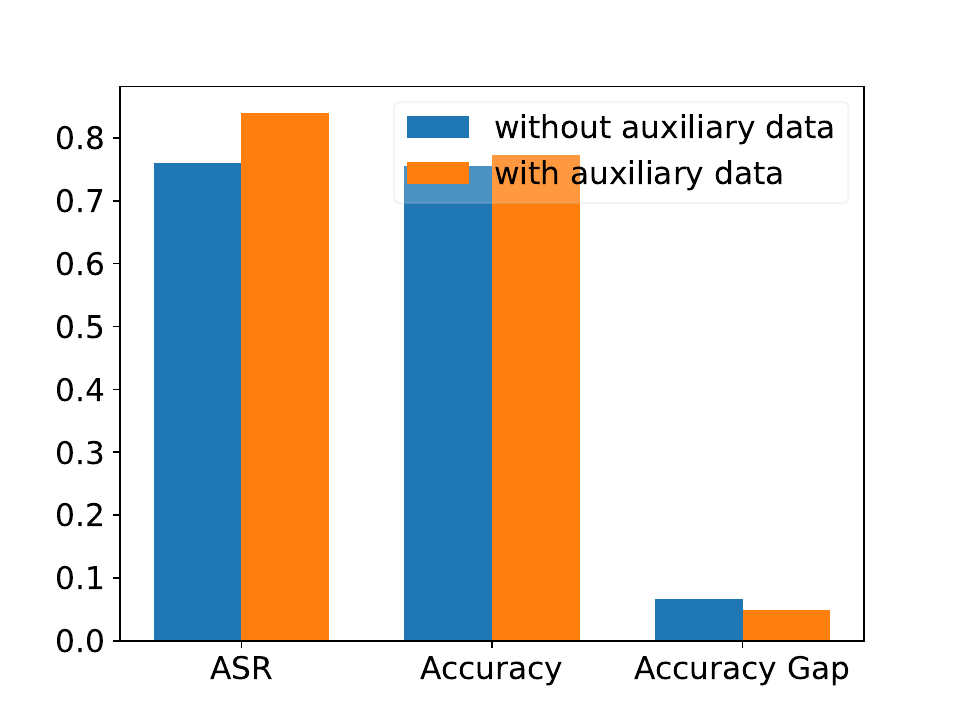}}
		\centerline{(a)}
	\end{minipage}
	\begin{minipage}{0.34\linewidth}
		\vspace{3pt}
		\centerline{\includegraphics[width=\textwidth]{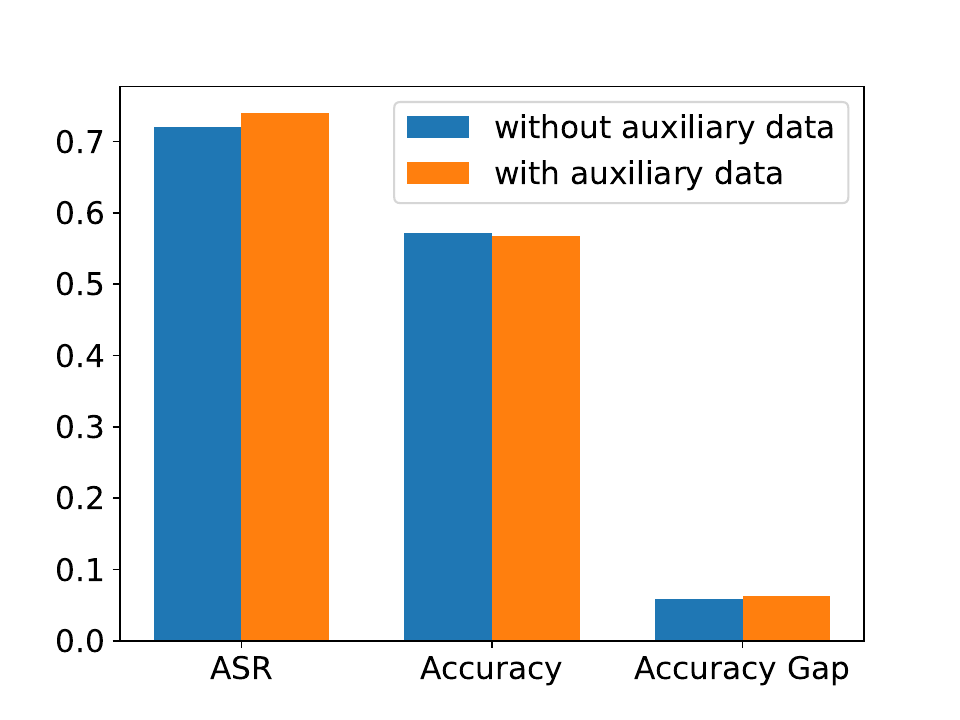}}
		\centerline{(b)}
	\end{minipage}
	\begin{minipage}{0.34\linewidth}
		\vspace{3pt}
		\centerline{\includegraphics[width=\textwidth]{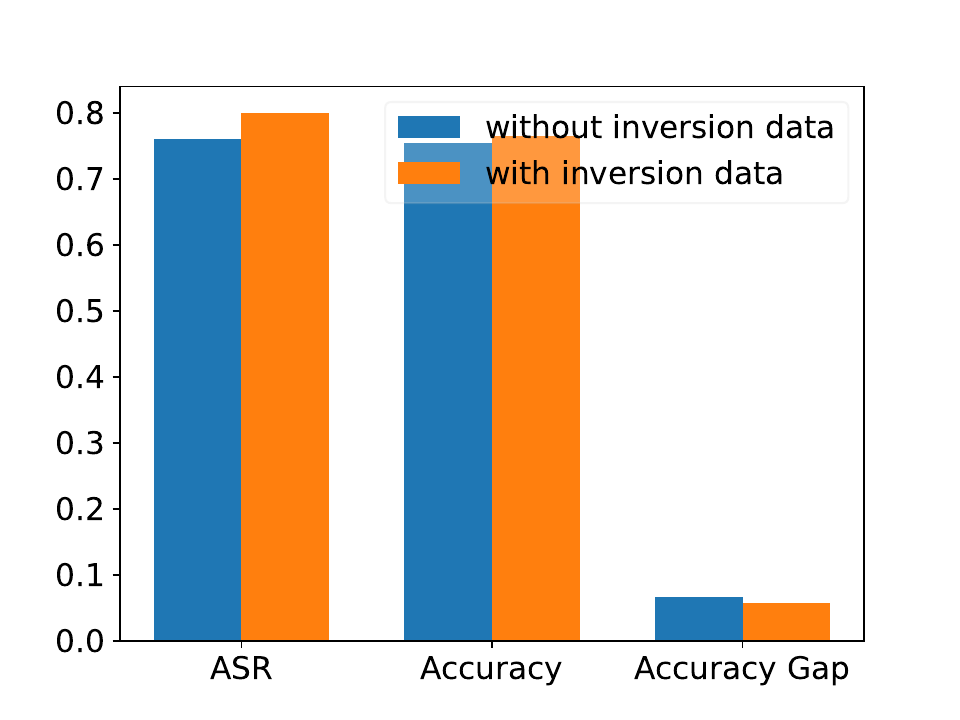}}
		\centerline{(c)}
	\end{minipage}
	\caption{Performance comparison of the proposed attack with and without augmented data. (a) and (b) show the performance of the proposed attack with and without auxiliary dataset on FaceScrub and Cifar100 datasets respectively. (c) demonstrates the performance comparison with and without inversion data on FaceScrub dataset.}
	\label{fig:auxiliary}
\end{figure*}

Figure~\ref{fig:auxiliary}(a) showcases the performance of the attack utilizing auxiliary data on the FaceScrub dataset. The FaceScrub dataset was divided into three parts: data for training the victim model, data for injecting the backdoor, and the auxiliary data. Notably, the auxiliary dataset does not include any data points used for model training or backdoor injection. As depicted in Figure~\ref{fig:auxiliary}(a), the attack success rate and the accuracy on benign samples both exhibit improvements when the auxiliary data is incorporated into the FaceScrub dataset. Additionally, Figure~\ref{fig:auxiliary}(b) presents a comparison of the attack performance with and without the auxiliary dataset on the Cifar100 dataset. Unlike the FaceScrub dataset, the proposed attack's performance shows minimal improvement when utilizing the auxiliary data on the Cifar100 dataset. This disparity can be attributed to the fact that the FaceScrub dataset consists of facial images of individuals sharing many similarities, thereby aligning well with the distribution of the auxiliary dataset. Consequently, the auxiliary dataset significantly aids in knowledge distillation, leading to notable performance enhancements. Conversely, the Cifar100 dataset encompasses images from different classes with varying distributions, resulting in a pronounced divergence between the distribution of the training data and that of the auxiliary dataset. Therefore, in the case of the Cifar100 dataset, the auxiliary dataset offers limited assistance in improving the attack performance.
\subsection{Backdoor attack with model inversion}
\begin{figure}
    \centering
    \includegraphics[width=\linewidth]{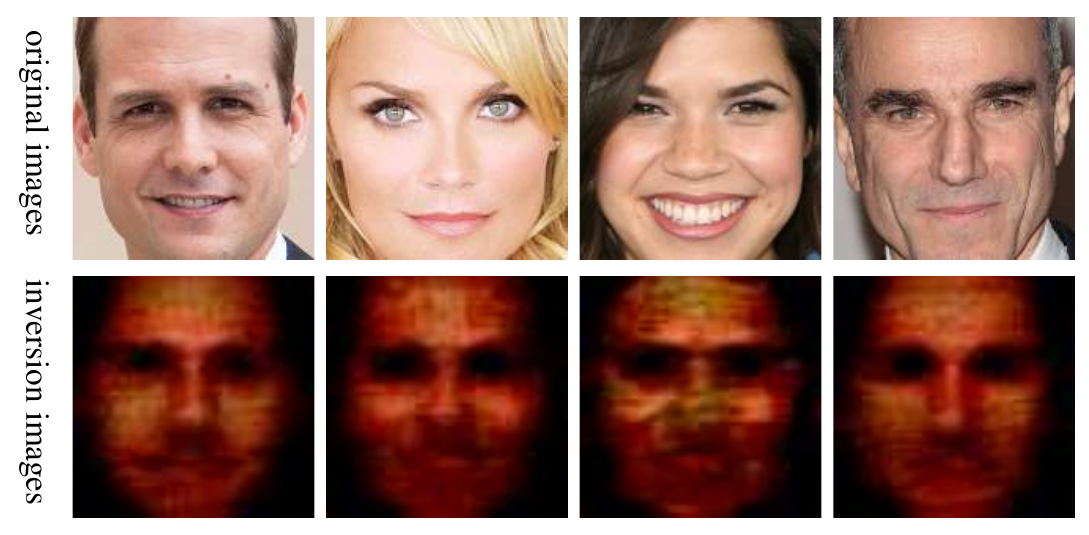}
    \caption{The original training images and the corresponding inversion images.}
    \label{fig:inversion}
\end{figure}

We also investigated the effectiveness of the proposed attack by incorporating model inversion techniques~\cite{fredrikson2015model}. In certain scenarios, it can be challenging to collect an auxiliary dataset with a distribution similar to the training data. Therefore, we explored the potential of model inversion techniques to aid the proposed attack. Model inversion techniques aim to reconstruct the training data used for training a DNN model, thereby potentially leading to privacy leakage. However, from another perspective, model inversion can generate images with representative features of the target class. This characteristic allows us to utilize inversion images to maintain the performance of the infected model on benign samples. The model inversion process is illustrated in Algorithm~\ref{alg:inversion}, where the initialized image $\bar{\bm{x}}$ is optimized to minimize the cross-entropy loss with respect to the target label $t$, resulting in the generation of images with representative features associated with label $t$. Additionally, we employed total variation loss to denoise the inversion images, which is defined as follows:
\begin{equation}
    \bm{\mathrm{TV}}\left( \bm{x}\right) = \sum_{i,j}\sqrt{\left(\bm{x}_{i+1,j}-\bm{x}_{i,j}\right)^2 + \left(\bm{x}_{i,j+1} - \bm{x}_{i,j} \right)^2}. \nonumber
\end{equation}
The inversion images and the corresponding ground-truth images are depicted in Figure~\ref{fig:inversion}. It is important to note that, in most cases, the inversion image does not closely resemble the original training data. However, despite this visual dissimilarity, the inversion image fulfills the same purpose as real images when training the neural network. If we train the model using both the original training set and the inversion data, the resulting neural networks achieve comparable accuracy.
\begin{algorithm}[h]
    \caption{Model Inversion Process}
    \label{alg:inversion}
    \SetKwInOut{Input}{Input}
    \SetKwInOut{Output}{Output} 
    \Input{The pre-trained model $M_{\bm{\theta}}$, the label of the target class $t$, the number of classes $K$, the number of inversion samples per class $m$, and the learning rate $\eta$ for inversion process.}
    \Output{The inversion samples $D_{inv}=\left\{ \left( \bar{\bm{x}}_1, \bar{y}_1 \right), \cdots,  \left( \bar{\bm{x}}_{km}, \bar{y}_{km}\right)\right\}$}
    \BlankLine
    $D_{inv} \leftarrow \left\{ \right\}$ \\
    \For{$i$ in $\mathrm{range}\left(n \right)$}{
        \For{$k$ in $\mathrm{range}\left( K \right)$}{
            $\bar{\bm{x}} \leftarrow \mathrm{init}\left( \right)$\\
            \For{iter in num\_iters}{
                $\bm{y} = M_{\bm{\theta}}\left(\bar{\bm{x}} \right)$\\
                $\mathcal{L}_{inv} = \bm{\mathrm{cross\_entropy}}\left(\bm{y}, t \right) + \mu \cdot \bm{\mathrm{TV}}\left(\bar{\bm{x}} \right)$\\
                $\bar{\bm{x}} \leftarrow \bar{\bm{x}} - \eta \frac{\partial \mathcal{L}_{inv}}{\partial \bar{\bm{x}}}$ \\
            }
            $D_{inv}.\bm{\mathrm{append}}\left( \left(\bar{\bm{x}}, k\right)\right)$ \\
        }
    }
    \textbf{return} $D_{inv}$
\end{algorithm}

Following the model inversion process, the inversion images retain certain representative features from the model's training data, enabling the attacker to utilize them for sustaining the performance of the infected model on benign samples. To achieve this objective, we employ the inversion data to calculate the distillation losses. The loss function can be formulated as follows:
\begin{equation}
\resizebox{0.91\linewidth}{!}{$
    \displaystyle
    \begin{aligned}
        \mathop{\min}_{\hat{\bm{\theta}}} L &= \sum_{\hat{\bm{x}} \in D_{Adv}}\mathcal{L}\left( M_{\hat{\bm{\theta}}}\left( \hat{\bm{x}}\right), \hat{y}\right) + \alpha \sum_{\bar{\bm{x}} \in D_{inv}} L_D\left( M_{\bm{\theta}}\left( \bar{\bm{x}}\right),  M_{\hat{\bm{\theta}}}\left( \bar{\bm{x}}\right)\right) \\
        &+ \beta \sum_{\bar{\bm{x}} \in D_{inv}}L_{AD}\left( M_{\bm{\theta}}\left( \bar{\bm{x}}\right), M_{\hat{\bm{\theta}}}\left( \bar{\bm{x}}\right)\right) + L_{EWC}. \nonumber
    \end{aligned}
$}
\end{equation}
As depicted in Figure~\ref{fig:auxiliary}(c), the utilization of inversion data to enhance the attack on the FaceScrub dataset has resulted in further improvements in the performance of the proposed attack. Notably, the generated data obtained through model inversion does not rely on any external knowledge. This characteristic makes it a powerful technique, particularly in low-resource scenarios where access to additional data is limited, apart from the malicious data owned by the attacker. Consequently, the inclusion of model inversion amplifies the harmfulness of the proposed attack.

\section{Conslusion} \label{sec:conclusion}
In this paper, we proposed a novel trigger-free backdoor attack on neural networks via a data-free manner. Specifically, we design a novel fine-tuning approach that incorporates the concept of malicious data into the concept of the attacker-specified class, resulting the misclassification of trigger-free malicious data into the attacker-specified class. Therefore, it is hard for the victim to inspect or detect the proposed backdoor attack via trigger-based defense methods, hence is more stealthy and malicious compared with existing attacks. Furthermore, the proposed attack does not need any training data of the benign model for backdoor injection to maintain the performance on benign samples, which is more practical for real scenarios. We leverage knowledge and attention distillation methods to constrain the consistency between the outputs of the infected and benign models on benign samples, and EWC method is adopted to preserve important knowledge of the benign model. Experiments show that the proposed model is capable to achieve high attack success rate while preserve the performance on benign samples.

In the future work, we plan to design defense methods for the proposed attack. We plan to detect the injected backdoor by analyzing the parameters of the model via a EWC approach. In addition, we also plan to apply the proposed attack to other tasks or models, such as LLM and text-to-image models.

\bibliographystyle{IEEEtranS}
\bibliography{ref}

\end{document}